# Thermal imaging by utilization of $CsPbBr_3$ quantum dot photoluminescence

**Alireza Jalouli[1], Nathan Giannini-Hutchin[1,2], Alexander R. Albrecht[1]**

[1] *Department of Physics & Astronomy, University of New Mexico, Albuquerque, NM 87131, USA*

[2] *currently at Sandia National Laboratories, Albuquerque, NM 87123, USA*

alex2@unm.edu

ajalouli@unm.edu

**Abstract:** The temperature-dependent photoluminescence of perovskite quantum dots ($CsPbBr_3$) in the visible band, is analyzed to evaluate their suitability for use in thermometry. A differential measurement of the photoluminescence can be used to estimate the surface temperature. Thermal imaging is demonstrated by using the Bayer-pattern of a cost-effective consumer-grade digital camera to determine the spectral shift. The temperature change of traces on a printed circuit board are visualized as proof of principle. This technique promises a novel approach for thermal imaging of arbitrary samples with optical resolution (wavelength of ~ 500 nm), instead of typical black-body wavelengths (~ 10 μm). Premature degradation of the quantum dots has been mitigated by embedding them in a poly-methyl methacrylate (PMMA) film, which can then be applied to arbitrary surfaces.

## 1. Introduction

Non-contact or remote temperature measurements are often performed using pyrometers or thermal cameras that measure the amount of black body radiation emitted by an object, typically in the mid- and far infrared (IR) part of the spectrum [1−3]. In the case of the ubiquitous uncooled microbolometer-based devices working in the 8 μm – 14 μm wavelength range, this requires specialized optics, typically made from Germanium, as fused silica and similar glasses are not transparent in this spectral region [4,5]. In addition, knowledge of the emissivity of the imaged object is also needed to accurately measure the temperature, with low-emissivity surfaces like polished metals being especially hard to calibrate [6].

For biological applications, the luminescence from organic dyes or semiconductor quantum dots has been used as a temperature sensor, typically by measuring the intensity of the luminescence [7 − 8]. While this technique can be very sensitive; it is susceptible to errors due to non-uniform intensity distribution or photo bleaching in the case of dyes. Moreover, more common Cd-based quantum dots are of high health risk and recent trends in public safety are toward Pb-free approaches [9]. Here, we investigate the use of the luminescence wavelength and lineshape, specifically of $CsPbBr_3$ perovskite quantum dots (QDs)[9 − 11].

While spectral information has been used for thermometry and investigations of cooling in rare-earth doped systems and bulk semiconductors [12 − 17], these emission lines are fixed at various positions around VIS-NIR band. In contrast, the size-dependent emission wavelength of QDs can be tailored over a large range, covering most of the visible and near-IR spectrum [16,17]. QD-based layers can be applied to arbitrary surfaces, enabling devices to capture luminescence and generate thermal images of arbitrary surfaces temperature, including biological systems, in a cost-effective and reliable manner [15]. One drawback of perovskite QD-based thermometry has been long-term fluorescence degradation resulting from oxidation of the solid-state emitter when in contact with air [10]. To enable robust QD coatings and ensure long-term operation, we passivated QDs by encapsulating them in PMMA [18 −19]. Photoluminescence is excited using a laser diode or LED near 405 nm, which results in high pump absorption, but also avoids spectral overlap with the QD emission near 500 nm. We apply a technique of differential luminescence thermometry (DLT) [20 − 23], correlating sample temperature with the spectral shift as measured using a silicon CCD-line spectrometer. To showcase the ease of use of this method, we demonstrate thermal imaging using only the Bayer pattern on the sensor of a consumer-level DSLR camera [24].

## 2. Sample preparation

Given the goal of simple spectroscopy and imaging, high efficiency luminescence in the visible spectrum was desired. We used ~ 7-nm CsPbBr$_3$ QDs in toluene with a concentration of approximately 17-25 μM, synthesized at the University of Notre Dame [10] to produce peak emission wavelength around 509 nm. After temperature cycling from 300 to 360 K, the PL peak of the bare CsPbBr$_3$ QDs permanently shifted from its initial state, which would render the QDs unusable for temperature measurement. We attribute this to irreversible oxidation of the QDs [25]. To prevent oxidation and help attach the QDs to a surface after the solvent evaporates, we disperse the QDs in a PMMA polymer matrix at a 1:10 ratio (QD/toluene solution to PMMA), which is then sprayed onto 2.2 × 2.2 cm² cover slips using an airbrush to ensure uniform coating. The solvent is evaporated by placing the samples on a hot plate at approximately 340 K for a few minutes.

To assess the quality of the coated QD film, photoluminescence (PL) spectra were recorded. A laser diode (405 nm) with an output power of ~ 8 mW was used for excitation. Measurements were taken at multiple points along the diagonal of the sample. At each point, the sample was illuminated by the laser diode, and the PL light was collected by a convex lens of 20 mm focal length and directed into a 200 μm core diameter multi-mode fiber connected to an Ocean Optics USB2000 optical spectrometer. A Savitzky-Golay filter in MATLAB (polynomial order m = 3, frame length $l_f$ = 15) is applied to the PL spectra to lower the noise, being similar to near-neighbor averaging. The spectra are then normalized to their maximum intensity, as shown in Fig. 1(a). In the raw PL spectra data set, the variation in PL peak positions remains within a few nanometers around an average of 509 nm [Fig. 1(b)], attributed to differences in film thickness and the size variability of the QDs.

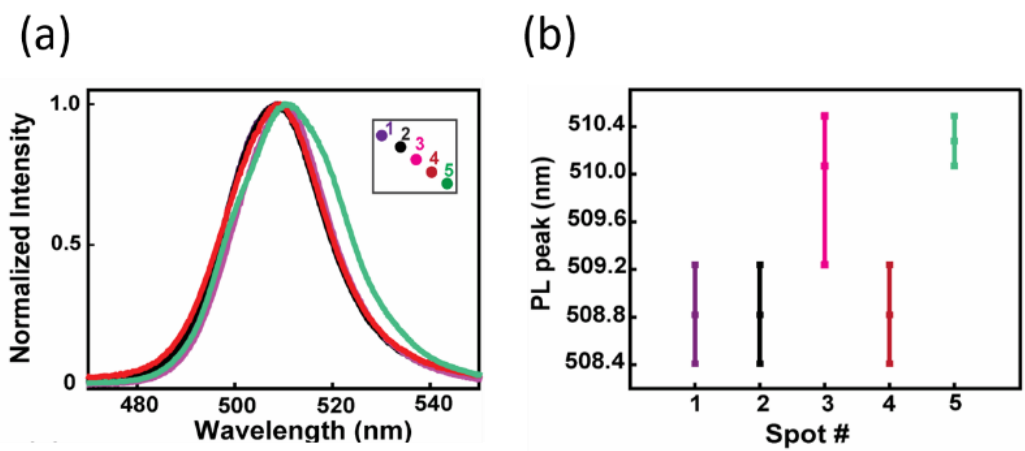

Fig.1 (a) PL spectra from the QD film were measured at various spots diagonally on the substrate. (b) PL peak wavelength at each spot on the film.

## 3. Temperature-dependent PL measurements

The goal of this work is to utilize the semiconductor QDs for non-contact temperature measurements and thermal imaging. Using a similar setup as in the previous section, but with the sample placed on a heating stage, PL spectra at various temperatures are collected, as shown in Fig. 2(a). A precise method to characterize temperature dependence is differential luminescence thermometry (DLT) [23], where the PL spectrum at room temperature (295 K) is used as a reference. All PL spectra at various temperatures are normalized to their peak intensity, and the room temperature PL spectrum is then subtracted from each normalized spectrum to assess the temperature-dependent changes [Fig. 2(b)]. It is evident in Fig. 2(b) that as the temperature increases, the PL peak redshifts and slightly broadens. This results in the increasing amplitude of the spectrally-integrated DLT signal, as shown in Fig. 2(c), where the total magnitude of the area under the curves in Fig. 2(b) is plotted as a function of temperature. This can be used to calibrate the DLT behavior, providing a surface temperature measurement. In Fig. 2(d), we illustrate a 12 nm redshift in the PL peak wavelength as the sample temperature is increased from 295 K to 360 K for three different samples. This temperature-induced redshift does not follow a linear relationship and increases significantly at temperatures higher than T = 335 K. This nonlinearity is also reflected in the full width at half maximum (FWHM) of the PL spectra, shown in Fig. 2(e), indicating a potential phase transition in the crystal structure of the QD nanocrystals around 335 K [24].

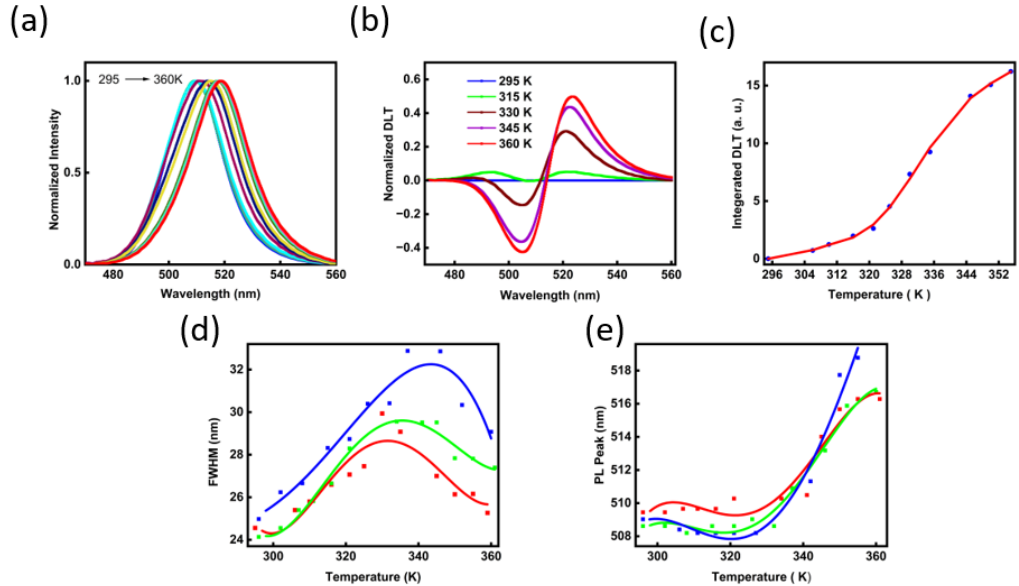

Fig. 2. (a), (b) PL and DLT curves of the $CsPbBr_3$ QDs at different temperatures ranging from 295 to 360 K. (c) The data points represent the integrated area of the curves in (b) and are fit by the solid red curve (4th-order polynomial). (d) and (e) PL peak and FWHM at different temperatures for three different samples are shown as dots, while the solid curves represent 4th-order polynomial fits.

## 4. Thermal imaging

As demonstrated in the previous section, differential analysis of the measured PL spectrum is a sensitive method for detecting subtle spectral changes induced by temperature. This approach typically requires the use of high-resolution spectrometers and various optical filters for optimal performance [26]. However, here we demonstrate a practical method for measuring surface temperatures using a cost-effective commercial digital single-lens reflex (DSLR) camera with Bayer-filter instead of complex laboratory instruments. This approach leverages image processing techniques to analyze the intensity of the different color pixels and correlate these variations with temperature changes

In Fig. 3(c), the images captured by a DSLR camera show the coated film sample illuminated under a 385-nm LED. To quantify this change with respect to the coated film's temperature, the following ratio is calculated: $(B - R)/(B + R)$, where B and R represent the pixel counts for the blue and red-colored pixels, respectively. The rationale behind this choice of formula will be clarified later.

The DSLR camera captures these images in an 8-bit format, meaning each pixel is represented by three numbers ranging from 0 to 255, corresponding to the intensities of the blue, red, and green colors. Higher numbers indicate greater intensity of the respective color, and the combination of the R, G, and B values determines the color of the pixel. Figure 3(a) illustrates the sensitivity of a typical DSLR camera to different colors [24]. Notably, the emission spectrum of $CsPbBr_3$ quantum dots, centered at 509 nm and used in this study, lies approximately midway between the red and blue regions, where the camera sensitivity is balanced.

The average ratio $(B - R)/(B + R)$ was calculated at various spots on the sample at room temperature (295 K) and at elevated temperatures up to 355 K. The room temperature ratio was then subtracted from the ratios at higher temperatures [Fig. 3(b)]. Results indicate that the magnitude of this calculated value, termed $A$, is greater at higher temperatures than at room temperature (295 K). The following outlines the calculation for $A$.

$$A = \frac{B-R}{B+R} - \frac{B_o - R_o}{B_o + R_o} \qquad (1)$$

The first term is calculated at a higher temperature, while the second term, which is based on pixel counts $B_o$, $R_o$ calculated at room temperature (295 K), serves as the reference.

The dashed line in Fig. 3(b) represents a linear fit, approximating the correlation between $A$ and temperature. This relationship highlights why $A$ is a suitable parameter for thermal imaging, as the magnitude of $A$ increases linearly with surface temperature. Figure 3(d) presents the calculation of $A$ for each pixel across the images captured by the camera in Fig. 3(c). The increasing intensity of the red areas, representing the magnitude of $A$, clearly corresponds to higher temperatures.

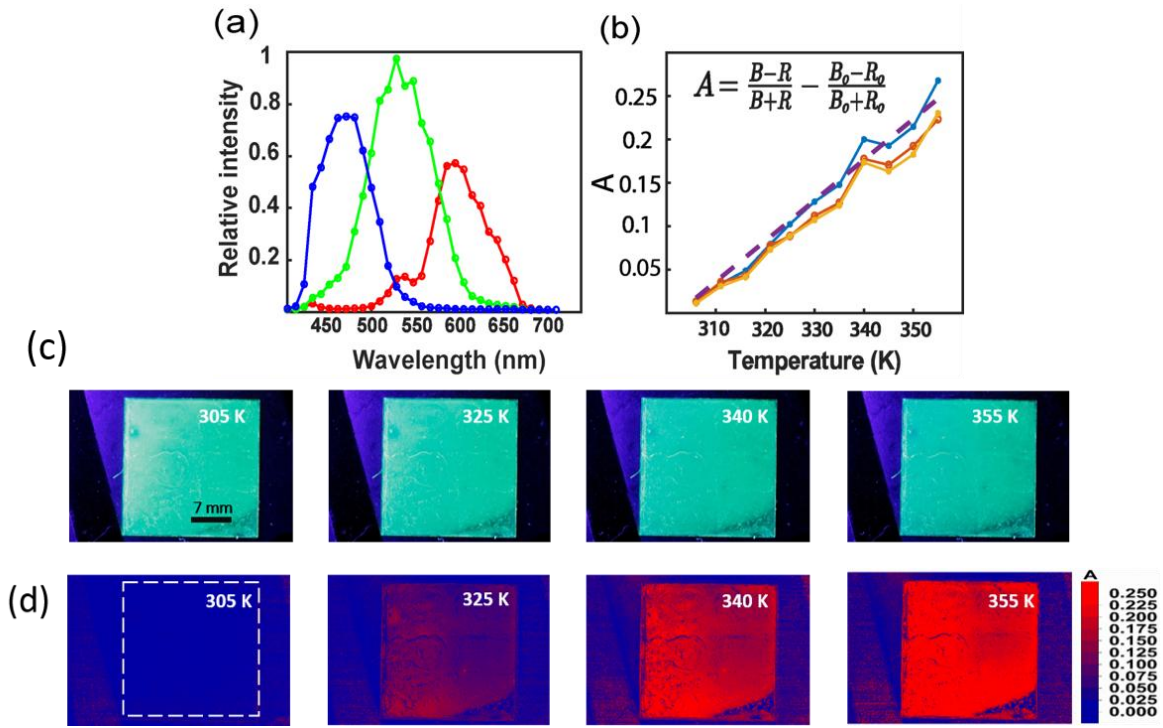

Fig. 3. (a) DSLR camera sensitivity diagram in the visible spectrum, showing the 3 different color filters of the Bayer pattern [27]. (b) Average ratio A at different temperatures. (c) Upper panel: photographs taken by DSLR at T = 305, 325, 345, and 355 K. (d) The images generated using eqn. (1) of the above photographs.

We believe this technique could be applied to a variety of applications for non-contact temperature sensing. As both the excitation and fluorescence wavelengths are in the visible spectrum, these measurements can be carried out using affordable, widely available light sources, optics, and detectors, instead of mid to far-IR thermal imaging equipment. Also, since the fluorescence of the QDs is utilized, knowledge of the thermal emissivity of the material to be measured is not needed for temperature measurement. More importantly, the spatial resolution achievable is limited by optical, rather than thermal wavelengths, potentially offering sub-micron resolution in microscopy or similar applications.

As proof of concept of our thermal imaging technique using a regular Bayer-pattern Silicon CCD camera, we coated the backside of a printed circuit board [Fig. 4(a)] with the QD-PMMA mixture. By applying a current to only one of the traces, we can selectively heat that one trace. Using the analysis presented in the previous section, we can create thermal images of the circuit board for varying currents, as shown in Fig. 4(c). As the applied current varies, the temperature of the trace's surface increases from room temperature to approximately 340 K

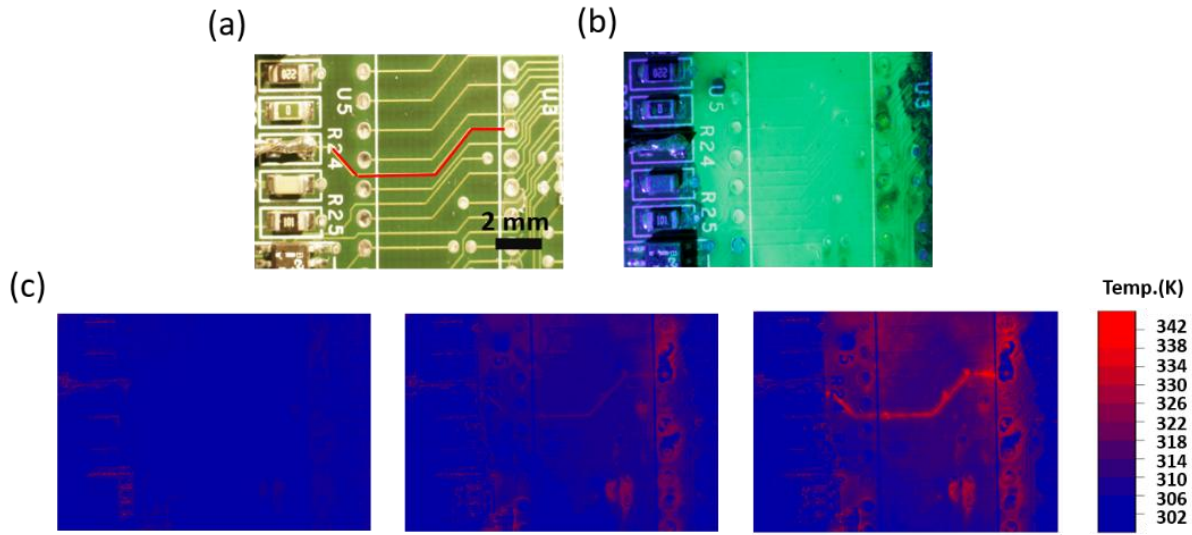

Fig. 4. (a) Visible light photograph of circuit board; a DC current can be applied to the trace highlighted in red. (b) The same circuit board under UV illumination, showing the area coated in QDs/PMMA. (c) Images calculated using eqn. [1] for currents of 0 A, 1.23 A, and 2.52 A, from left to right, respectively.

## 5. Conclusion

In conclusion, we introduced a simple and cost-effective method for capturing thermal images of various surfaces with fine features using basic commercial-grade digital cameras. The use of a removable film coating incorporating quantum dots (QD) within a Polymethyl-methacrylate (PMMA) matrix proves to be an effective solution. The correlation between the photoluminescence (PL) spectra of the coated film and the temperature provides a unique avenue for indirectly measuring surface temperature. The emission peak of $CsPbBr_3$ QDs around 500 nm suggests that a basic DSLR camera, with sufficient sensitivity within this wavelength range, can be utilized effectively. This 500 nm wavelength offers thermal imaging with enhanced spatial resolution compared to conventional thermal cameras, which operate in the domain of a few micrometers.

## 6. Acknowledgment

We express our gratitude to Prof. Masaru Kuno and Mr. Yang Ding for generously synthesizing the QDs in their laboratory at the University of Notre Dame. Special thanks to Dr. Francesca Cavallo from the University of New Mexico for providing the PMMA.
This work was conceived by and conducted under the supervision of Prof. Mansoor Sheik-Bahae, who passed away in 2023. The authors are extremely grateful for his guidance and support, he is sorely missed.
This research was in part funded by Air Force Office of Scientific Research under FA9550-16-1-0362(MURI).

**References**

1. Woodruff C, Dean SW, Pantoya ML. Comparison of pyrometry and thermography for thermal analysis of thermite reactions. *Appl Opt*. 2021;60(16):4976–4985. doi:10.1364/AO.423924

2. Lafargue-Tallet T, Vaucelle R, Caliot C, et al. Active thermo-reflectometry for absolute temperature measurement by infrared thermography on specular materials. *Scientific Reports*. 2022;12(1):7814. doi:10.1038/s41598-022-11616-8

3. Manullang MC, Lin YH, Lai SJ, Chou NK. Implementation of thermal camera for non-contact physiological measurement: A systematic review. *Sensors*. 2021;21(23):7777. doi:10.3390/s21237777

4. Kruse PW. *Uncooled Thermal Imaging: Arrays, Systems, and Applications*. Bellingham, WA: Society of Photo Optical Instrumentation Engineers (SPIE); 2001.

5. Torres A, Kosarev A, García Cruz ML, Ambrosio R. Uncooled micro-bolometer based on amorphous germanium film. *Journal of Non-Crystalline Solids*. 2003;329(1):179–183. doi:10.1016/j.jnoncrysol.2003.08.037

6. Jia Y, Liu D, Chen D, et al. Transparent dynamic infrared emissivity regulators. *Nature Communications*. 2023;14(1):5087. doi:10.1038/s41467-023-40902-w

7. Jaiswal JK, Simon SM. Potentials and pitfalls of fluorescent quantum dots for biological imaging. *Trends in Cell Biology*. 2004;14(9):497–504. doi:10.1016/j.tcb.2004.07.012

8. Barroso MM. Quantum dots in cell biology. *J Histochem Cytochem*. 2011;59(3):237–251. doi:10.1369/0022155411398487

9. Giroux MS, Zahra Z, Salawu OA, Burgess RM, Ho KT, Adeleye AS. Assessing the environmental effects related to quantum dot structure, function, synthesis and exposure. *Environ Sci: Nano*. 2022;9(3):867–910. doi:10.1039/D1EN00712B

10. Ding Y, Zhang Z, Toso S, et al. Mixed ligand passivation as the origin of near-unity emission quantum yields in CsPbBr3 nanocrystals. *J Am Chem Soc*. 2023;145(11):6362–6370. doi:10.1021/jacs.2c13527

11. Tang F, Su Z, Ye H, Zhu Y, Dai J, Xu S. Anomalous variable-temperature photoluminescence of CsPbBr3 perovskite quantum dots embedded into an organic solid. *Nanoscale*. 2019;11(43):20942–20948. doi:10.1039/C9NR07081H

12. Dirin DN, Cherniukh I, Yakunin S, Shynkarenko Y, Kovalenko MV. Solution-grown CsPbBr3 perovskite single crystals for photon detection. *Chem Mater*. 2016;28(23):8470–8474. doi:10.1021/acs.chemmater.6b04298

13. Seletskiy DV, Epstein R, Sheik-Bahae M. Laser cooling in solids: Advances and prospects. *Reports on Progress in Physics*. 2016;79(9):096401. doi:10.1088/0034-4885/79/9/096401

14. Clark AM, Miller NA, Williams A, et al. Cooling of bulk material by electron-tunneling refrigerators. *Applied Physics Letters*. 2005;86(17):173508. doi:10.1063/1.1914966

15. Dramićanin MD. Trends in luminescence thermometry. *Journal of Applied Physics*. 2020;128(4):040902. doi:10.1063/5.0014825

16. Rupper G, Kwong NH, Binder R. Large excitonic enhancement of optical refrigeration in semiconductors. *Phys Rev Lett*. 2006;97(11):117401. doi:10.1103/PhysRevLett.97.117401

17. Nguyen DH, Kim SH, Lee JS, Lee DS, Lee HS. Reaction-dependent optical behavior and theoretical perspectives of colloidal ZnSe quantum dots. *Scientific Reports*. 2024;14(1):13982. doi:10.1038/s41598-024-64995-5

18. Gao L, Quan LN, García de Arquer FP, et al. Efficient near-infrared light-emitting diodes based on quantum dots in layered perovskite. *Nature Photonics*. 2020;14(4):227–233. doi:10.1038/s41566-019-0577-1

19. Yuan F, Folpini G, Liu T, et al. Bright and stable near-infrared lead-free perovskite light-emitting diodes. *Nature Photonics*. 2024;18(2):170–176. doi:10.1038/s41566-023-01351-5

20. Lee SM, Moon CJ, Lim H, Lee Y, Choi MY, Bang J. Temperature-dependent photoluminescence of cesium lead halide perovskite quantum dots: Splitting of the photoluminescence peaks of CsPbBr3 and CsPb(Br/I)3 quantum dots at low temperature. *J Phys Chem C*. 2017;121(46):26054–26062. doi:10.1021/acs.jpcc.7b06301

21. Cheng P, Zhu Y, Shi J, et al. One-step solution deposited all-inorganic perovskite CsPbBr3 film for flexible resistive switching memories. *Applied Physics Letters*. 2019;115(22):223505. doi:10.1063/1.5120791

22. Seletskiy DV, Hasselbeck MP, Sheik-Bahae M, Epstein RI. Fast differential luminescence thermometry. In: *Advanced Optical and Mechanical Technologies in Telescopes and Instrumentation*. Vol. 7228. SPIE; 2009:72280K. doi:10.1117/12.810856

23. Kock J, Zhang M, Priante D, Albrecht AR, Volpi A, Sheik-Bahae M. Perovskite quantum-dot photoluminescence for non-contact thermometry and thermal imaging [poster presentation]. In: *Photonic Heat Engines: Science and Applications II*. Vol. 11298. SPIE; 2020:11298-30.

24. Huang D, Bo J, Zheng R, et al. Luminescence and stability enhancement of CsPbBr3 perovskite quantum dots through surface sacrificial coating. *Advanced Optical Materials*. 2021;9(16):2100474. doi:10.1002/adom.202100474

25. Imran M, Ijaz P, Goldoni L, et al. Simultaneous cationic and anionic ligand exchange for colloidally stable CsPbBr3 nanocrystals. *ACS Energy Lett*. 2019;4(4):819–824. doi:10.1021/acsenergylett.9b00140

26. Liu W, Zhang Y, Zhai W, et al. Temperature-dependent photoluminescence of ZnCuInS/ZnSe/ZnS quantum dots. *The Journal of Physical Chemistry C*. 2013;117(38):19288–19294. doi:10.1021/jp4024603

27. Solomatov G, Akkaynak D. Spectral sensitivity estimation without a camera. In: *2023 IEEE International Conference on Computational Photography (ICCP)*. IEEE; 2023:1–12. doi:10.1109/ICCP56744.2023.10233713